\documentclass[pra,balancelastpage,twocolumn,footinbib,superscriptaddress]{revtex4-1}
\usepackage{color,amsthm,amsmath,amsfonts,graphicx,bm}
\usepackage{bm}
\usepackage{color}
\usepackage{amsfonts}
\usepackage{amssymb}
\usepackage{amsmath}
\usepackage{epsfig}
\usepackage{supertabular}
\usepackage{graphicx}
\usepackage{enumerate}
\usepackage{multirow}
\usepackage{verbatim}
\usepackage{color}
\usepackage{subfigure}

\theoremstyle{plain}
\theoremstyle{definition}
\usepackage{array}
\usepackage{booktabs}
\usepackage{tabularx}
\usepackage{makecell}
\begin{document}

\title{Adiabatic pumping of topological corner states by coherent tunneling in a 2D SSH model}

\author{Yang Peng}
\affiliation{College of Physics and Electronic Engineering, Institute of Solid State Physics, Sichuan Normal University, Chengdu 610068, China}
\author{Rui-Shan Li}
\affiliation{College of Physics and Electronic Engineering, Institute of Solid State Physics, Sichuan Normal University, Chengdu 610068, China}
\author{Yan-Jue Lv}
\affiliation{College of Physics and Electronic Engineering, Institute of Solid State Physics, Sichuan Normal University, Chengdu 610068, China}
\author{Yi Zheng}
\email{Corresponding author: zhengyireal@sicnu.edu.cn}
\affiliation{College of Physics and Electronic Engineering, Institute of Solid State Physics,  Sichuan Normal University, Chengdu 610068, China}
\pacs{}

\begin{abstract}
The active manipulation of topologically protected states represents a pivotal frontier for quantum technologies, offering a unique confluence of topological robustness and precise quantum control. We propose an adiabatic pumping scheme for the long-range transfer of topological corner states in a two-dimensional Su-Schrieffer-Heeger model. The protocol utilizes a modular lattice architecture composed of four topologically distinct subblocks, enabling the modulation of a topological dark state by precise tuning of lattice couplings. This approach is based on coherent tunneling by adiabatic passage among topological corner and interface states. We establish a multi-level model for the adiabatic pumping that provides an accurate description of the underlying mechanism. In comparison with a sequential two-stage Thouless pumping, our protocol offers superior performance in both transfer fidelity and efficiency. 

\end{abstract}

\maketitle

\section{Introduction}\label{intro}
Topological phases of matter have emerged as a cornerstone of modern condensed matter physics, revealing rich phenomena that challenge conventional notions of order and symmetry \cite{thouless1982quantized, schnyder2008classification, hasan2010colloquium, ryu2010topological, qi2011topological, bansil2016colloquium}. Prototype models, including the Su-Schrieffer-Heeger (SSH) model \cite{su1979solitons}, the Creutz ladder \cite{creutz1999end} and the Haldane model \cite{haldane1988model}, have provided instructive paradigms for exploring topological invariants, robust edge states and quantized transport \cite{delplace2011zak, grusdt2013topological, nakajima2016topological, sun2017quantum, junemann2017exploring, cooper2019topological}. One of the most prominent manifestations of topological phenomena is Thouless pumping \cite{thouless1983quantization, ke2017multiparticle, fedorova2020observation, citro2023thouless, lv2025exploring}, which describes adiabatic charge transport through periodic modulation of a one-dimensional (1D) system. The transported charge over a modulation cycle is quantized to the Chern number of filled band. This gives rise to an experimentally feasible approach to identify topological phases. As exemplified by the Rice-Mele scheme \cite{rice1982elementary}, such a quantum pump reveals the connection between the Zak phase for an SSH model and the two-dimensional (2D) topological invariant Chern number \cite{xiao2010berry, wang2013topological}. The global properties in topology endow the quantized pumping with remarkable robustness against local disorder and perturbations \cite{niu1984quantised, kraus2012topological, lohse2016thouless}. This intrinsic stability is particularly significant for high-fidelity quantum state control and robust excitation transfer \cite{xu2017topological, longhi2019topological, zurita2020topology, minguzzi2022topological}. Recent experiments have also demonstrated the implementation of such a robust quantum pump in the contexts of semiconductor quantum dots \cite{connolly2013gigahertz, kandel2021adiabatic}, cold atom systems \cite{lohse2016thouless, nakajima2016topological} and photonic waveguide arrays \cite{kraus2012topological}, and tunable electric circuits \cite{yatsugi2022observation}. 

Topological feature of the bulk band structure dictates the presence of localized states at the edges or interfaces, as a result of the bulk-boundary correspondence \cite{ryu2002topological}. The manipulation of edge states, including topological pumping, has garnered significant interest due to the potential applications in fault-tolerant quantum computing \cite{nayak2008non}, engineering of entangled topological states \cite{gorg2019realization, de2019observation}, as well as spintronics \cite{mellnik2014spin, han2018quantum}. Recent studies on topological pumping  of edge states have introduced an efficient scheme based on coherent tunneling by adiabatic passage (CTAP) \cite{longhi2019topological}. In such a protocol, an edge state can be pumped in a long chain from one boundary to another with the assistance of a topological protected interface state. This aligns with the spirit of stimulated Raman adiabatic passage (STIRAP) for internal state transfer, well known in atomic, molecular and optical (AMO) physics. In contrast to Thouless pumping schemes, CTAP and STIRAP protocols achieve coherent quantum state transfer by guiding the system through a dark state \cite{gaubatz1990population, bergmann1998coherent, menchon2016spatial, vitanov2017stimulated, leroux2018non}. This mechanism provides inherent robustness against pulse errors and decoherence from the intermediate state. While celebrated for robust population transfer in diverse systems, the application of these dark-state protocols to the control of higher-order topological states, such as corner modes, remains largely unexplored. 

Higher-order topology extends conventional (first-order) topological phases in a profound way, shifting the focus from bulk and edge states to localized corner or higher-dimensional vertex states. The idea is to generalize the notion of quantized electric dipole moments (characterized by the Zak phase) to quantized quadrupole moments \cite{benalcazar2017quantized, benalcazar2017electric, li2020topological, serra2018observation, yang2020type, vergara2024emerging, vergara2024emerging}, giving rise to topologically protected zero-energy corner states. The 2D SSH model provides a minimal candidate for realizing such higher-order topological phases \cite{xie2018second, obana2019topological, kim2020topological,  yang2021robust, li2022topological, wei2023topological}. Specifically, a direct 2D generalization of the SSH model consists of coupled 1D SSH chains arranged with alternating interchain tunneling amplitudes, leading to a lattice with staggered hopping strengths. However, zero-energy corner modes are located in the bulk continuum, as has been demonstrated in the photonic platform \cite{mittal2019photonic}. A synthetic $\pi$ flux per plaquette, which opens up a band gap at zero energy, has been proposed \cite{benalcazar2017electric}. This construction, also known as Benalcazar-Bernevig-Hughes (BBH) model, inherits and extends the essential topological characteristics of its 1D counterpart. The corner states offer spatially localized zero-dimensional modes that are promising candidates for hosting quantum bits or acting as nodes in a quantum network \cite{lu2014topological, el2019corner, deng2024high}. The unique spatial confinement also allows for additional pathways in excitation transfer \cite{benalcazar2022higher}. Given these features, it is of theoretical and practical interest to explore dynamic control mechanisms, such as adiabatic pumping, for corner states. As inspired by the CTAP protocol for edge states in a 1D SSH chain \cite{longhi2019topological}, in this work, we present a scheme for coherent corner state transfer via adiabatic passage. We introduce a superlattice combined by four 2D SSH models with distinct parameter sets of hopping rates. This allows for coherent tunneling between topological corner states and interface (corner) states existing on the vertical and horizontal junctures. We show that CTAP provides an efficient pumping scheme for corner state transfer, since the energy of localized corner modes remains in the band gap, which is in contrast to the Rice-Mele scheme in Thouless pumping. 

This paper is organized as follows. Section \ref{sec2} introduces the extended 2D SSH model and its topological corner and interface states. In Sec. \ref{sec3}, we present the adiabatic scheme based on the combination of topological corner state pumping and CTAP, giving rise to an efficient excitation transfer with high fidelity. An effective multi-level model within the topological states subspace is established to describe the transport process. This protocol is compared with a Thouless pumping scheme in Sec. \ref{sec4}. A summary is included in Sec. \ref{sec5}.

\begin{figure}[tbp]
	\includegraphics[width=8.6cm]{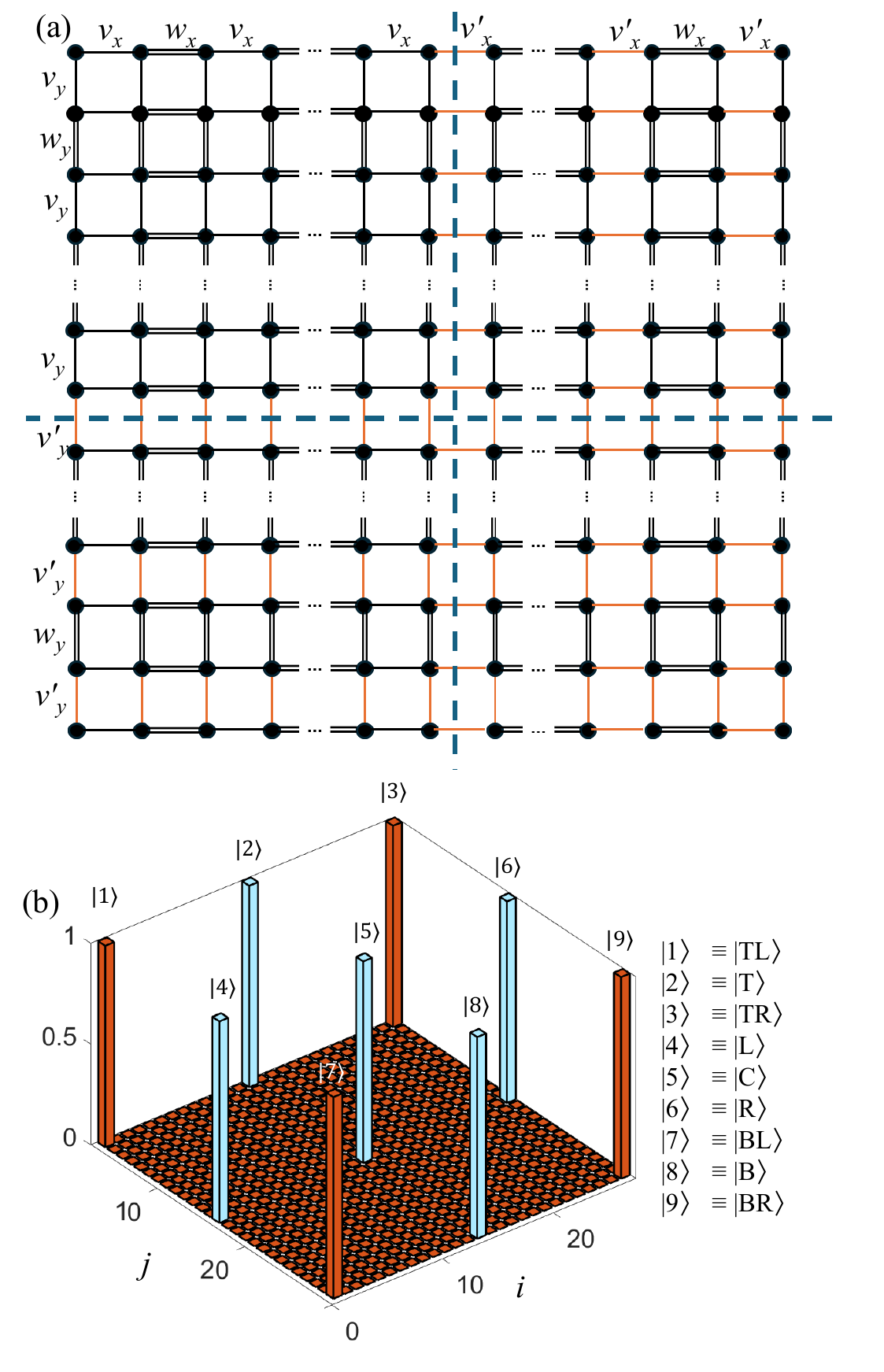}
\caption{(a) Schematic illustration of the 2D tight-binding model with interfaces labeled by dashed lines. The system is constructed by connecting four 2D SSH models. The corresponding hopping rates satisfy $w_{x(y)}>v_{x(y)},v_{x(y)}'$, leading to the existence of nine topological corner states. (b) Distributions of the nine degenerate topological corner modes in the flat-band limit $v_{x(y)}=v_{x(y)}'=0$. }
\end{figure}
\begin{table*}
  \centering
  \caption{topological corner states and interface corner states}
  \resizebox{0.75\textwidth}{!}{
  \begin{tabular}{l | c}
  \toprule[1pt]
  \midrule[0.5pt]
  \thead{state} & \thead{distribution (the summation runs over integers $s$, $s'\geq0$)} \\
		\midrule[0.8pt]
        \thead{$|TL\rangle$} & \thead{$\mathcal N_{tl}\sum_{s,s'}M_x^sM_y^{s'}\hat c_{2s+1,2s'+1}^\dag|0\rangle$}  \\
	    \thead{$|T\rangle$} & \thead{$\mathcal N_{t}\sum_{s,s'}\left[M_x^sM_y^{s'}\hat c_{L-2s,2s'+1}^\dag+N_x^{s+1}M_y^{s'}\hat c_{L+2(s+1),2s'+1}^\dag\right]|0\rangle$} \\
	    \thead{$|TR\rangle$} & \thead{$\mathcal N_{tr}\sum_{s,s'}N_x^sM_y^{s'}\hat c_{2L-1-2s,2s'+1}^\dag|0\rangle$ }\\
		\thead{$|L\rangle$} & \thead{$\mathcal N_{l}\sum_{s,s'}\left[M_x^sM_y^{s'}\hat c_{2s+1,L-2s'}^\dag+M_x^{s}N_y^{s'+1}\hat c_{2s+1,L+2(s'+1)}^\dag\right]|0\rangle$} \\
		\thead{$|C\rangle$} & \thead{$\mathcal N_{c}\sum_{s,s'}\left[\right.M_x^sM_y^{s'}\hat c_{L-2s,L-2s'}^\dag+N_x^{s+1}M_y^{s'}\hat c_{L+2(s+1),L-2s'}^\dag+M_x^sN_y^{s'+1}\hat c_{L-2s,L+2(s'+1)}^\dag$\\$+ N_x^{s+1}N_y^{s'+1}\hat c_{L+2(s+1),L+2(s'+1)}^\dag\left.\right]|0\rangle$ } \\
		\thead{$|R\rangle$} & \thead{$\mathcal N_{r}\sum_{s,s'}\left[N_x^sM_y^{s'}\hat c_{2L-1-2s,L-2s'}^\dag+N_x^{s}N_y^{s'+1}\hat c_{2L-1-2s,L+2(s'+1)}^\dag\right]|0\rangle$} \\
		\thead{$|BL\rangle$} & \thead{$\mathcal N_{bl}\sum_{s,s'}M_x^sN_y^{s'}\hat c_{2s+1,2L-1-2s'}^\dag|0\rangle$} \\
		\thead{$|B\rangle$} & \thead{$\mathcal N_{b}\sum_{s,s'}\left[M_x^sN_y^{s'}\hat c_{L-2s,2L-1-2s'}^\dag+N_x^{s+1}N_y^{s'}\hat c_{L+2(s+1),2L-1-2s'}^\dag\right]|0\rangle$} \\
		\thead{$|BR\rangle$} & \thead{$\mathcal N_{br}\sum_{s,s'}N_x^sN_y^{s'}\hat c_{2L-1-2s,2L-1-2s'}^\dag|0\rangle$} \\
		\midrule[0.5pt]
		\bottomrule[1pt]
  \end{tabular}}\label{table1}
\end{table*}
\section{coupled 2D SSH model}\label{sec2}
We consider a tight-binding model formed by a $2\times 2$ array of smaller 2D SSH model, as schematically shown in Fig. 1(a). The borders between the four sub-blocks constitute one vertical and one horizontal interface. Each individual chain along $x$ or $y$ direction can be regarded as a 1D lattice composed of two SSH chains with different topological orders \cite{benalcazar2017electric}. The Hamiltonian of the system is written as
\begin{equation}\label{eq_H}
	\hat H = \sum_{i,j} \left[ J_{i,j}^x\hat c^\dag_{i+1,j}\hat c_{i,j} + (-1)^iJ_{i,j}^y \hat c^\dag_{i,j+1}\hat c_{i,j} + \text{h.c.} \right].
\end{equation}
Here $i,j\in \left[1,2L-1\right]$ with $L$ even. $\hat c^\dag_{i,j}$ is the creation operator for particles on site $(i,j)$. A $\pi$ flux per plaquette has been implemented by imposing a tunneling phase ($e^{i\pi}$) to each odd column. The bond dimerization along $\hat x$ and $\hat y$ directions are given by
\begin{eqnarray}
	J_{i,j}^{x(y)} = \begin{cases}
		v_{x(y)},& i\in \text{odd}\\
		w_{x(y)},& i\in \text{even}
	\end{cases},\ \text{for } i<L;\\
	J_{i,j}^{x(y)} = \begin{cases}
		w_{x(y)},& j\in \text{odd} \\
		v_{x(y)}',&  j\in \text{even}
	\end{cases},\ \text{for } i\geq L.
\end{eqnarray}
Each sub-block is formed by a unit cell with four sites. Taking the upper left sub-block as an example, the Bloch Hamiltonian of the bulk reads as
\begin{eqnarray}\label{eq_Hk}
	\hat H_k&=&(v_x+w_x\cos k_x)I\otimes\sigma_x + w_x\sin k_xI\otimes\sigma_y\nonumber\\
	&-&(v_y+w_y\cos k_y)\sigma_x\otimes\sigma_z + w_y\sin k_y\sigma_y\otimes\sigma_z.
\end{eqnarray}
Here $\sigma_a$ ($a=x,y,z$) are Pauli matrices and $I$ is the identity matrix. We have taken $\{\hat A_{s,s'},\hat B_{s,s'},\hat C_{s,s'},\hat D_{s,s'} \} = \{\hat c_{2s,2s'},\hat c_{2s+1,2s'},\hat c_{2s,2s'+1},\hat c_{2s+1,2s'+1} \}$ (with $s$ and $s'$ integers) as the corresponding operators for the four modes in a unit cell. The system preserves reflection symmetries $R_x:x\leftrightarrow -x$ and $R_y:y\leftrightarrow -y$, which are noncommutative due to the presence of the flux. Such an arrangement opens up a band gap and hence isolates the topological quadrupole corner states. We mention that under open boundary condition, this individual sub-block with lattice size $L\times L$ supports four corner states when $w_x>v_x$ and $w_y>v_y$. This can be seen by rewriting Eq. (\ref{eq_Hk}) as $\hat H_k=I\otimes \hat H_{SSH} + \hat H_{SSH}'\otimes \sigma_z$ with $\hat H_{SSH}$ and $\hat H_{SSH}'$ in the form of the Bloch Hamiltonians of a 1D SSH model. However, in the topological nontrivial phase with $w_x>v_x'$ and $w_y>v_y'$, the lower right $(L-1)\times (L-1)$ sub-block, when decoupled from the others, sustains a single corner state  which resides on the bottom right vertex. For the complete lattice model shown in Fig. 1(a), we set $w_x>v_x,v_x'$ and $w_y>v_y,v_y'$ to guarantee the existence of four corner states ($|TL\rangle,|TR\rangle,|BL\rangle,|BR\rangle$) on the vertexes, four interface corner states ($|T\rangle,|L\rangle,|R\rangle,|B\rangle$) at the midpoints of each edge, and one soliton state ($|C\rangle$) at the center of the lattice. In the thermodynamic limit, the nine topological states and their distributions are presented in table \ref{table1}, where $\mathcal{N}_{\cdots}$ represent the corresponding normalization constants and we have set 
\begin{equation}
	M_{x(y)} = -\frac{v_{x(y)}}{w_{x(y)}},\ \ N_{x(y)} = -\frac{v_{x(y)}'}{w_{x(y)}}.
\end{equation}
We mention that the corner states on the vertexes and those at the midpoints of the edges are exact eigenstates of Eq. (\ref{eq_H}) with zero eigenvalues for a semi-infinite system. The center state $|C\rangle$ is also the zero-energy eigenstate for an infinite system. For a finite-size lattice, these states hybridize as a result of generally small overlaps. The eigenvalues of the hybridized states deviate from zero. For $v_{x(y)}=v_{x(y)}'=0$, such topological states become fully localized at the corresponding sites, as illustrated in Fig. 1(b). Increasing the values of $v_{x(y)}$ and (or) $v_{x(y)}'$ extends the wavefunctions of these states and therefore enhances their couplings. 

\section{corner state pumping by adiabatic passage}\label{sec3}
\subsection{Topological pumping scheme}
\begin{figure*}[tbp]
	\includegraphics[width=17cm]{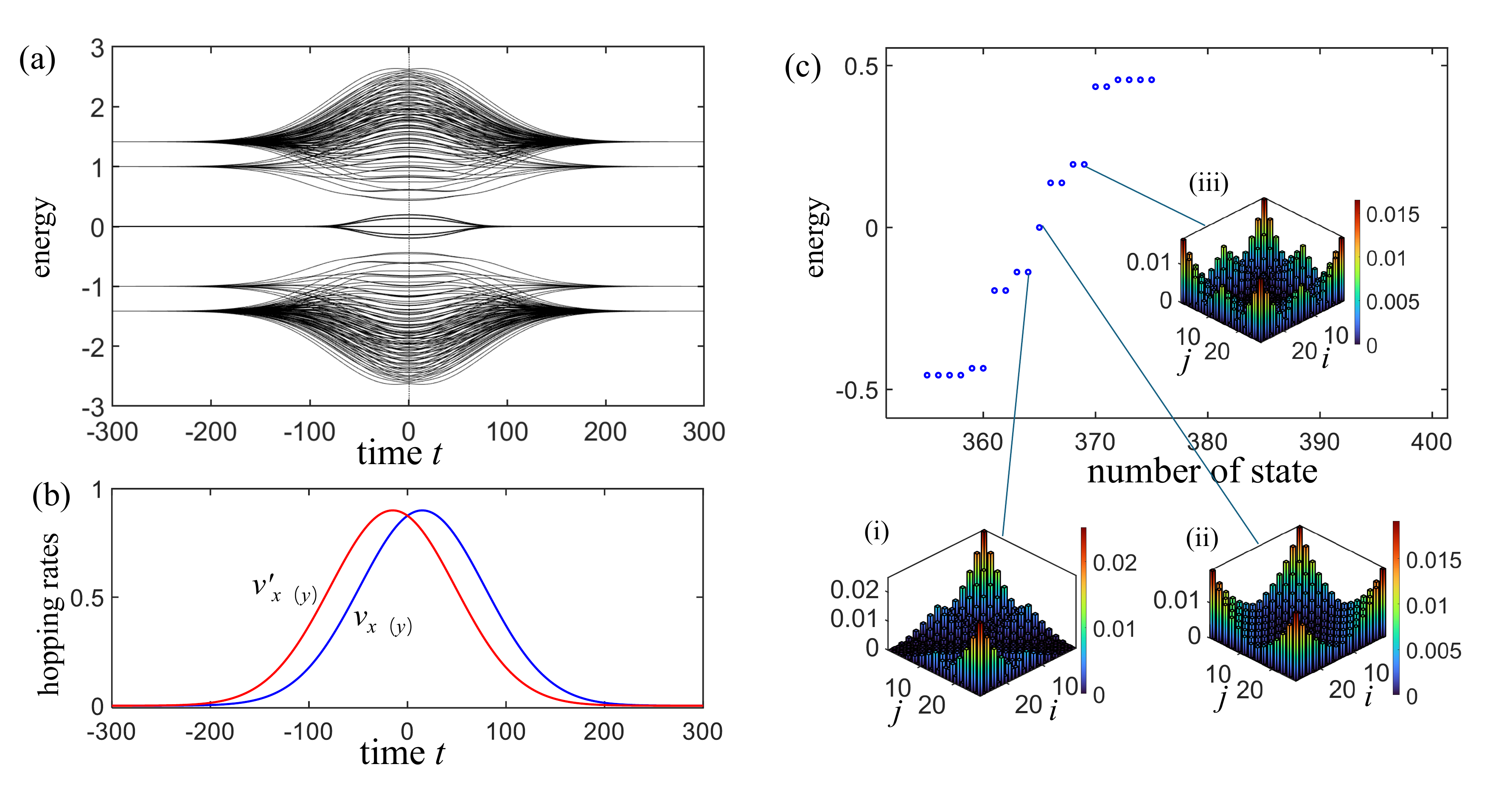}
\caption{
(a) Instantaneous energy spectrum of $H(t)$ with $2L - 1 = 27$. (b) Time dependence of the hopping amplitudes. We have set $v_x=v_y$, $v_x'=v_y'$, which follow  the bell-shaped function $\Omega(t) = \Omega_m \exp[-(t\pm\delta/2)^2/\lambda^2]$. The parameters are $\Omega_m = 0.9$, $\lambda = 150$, and $\delta = w/3$. Within the energy gap of the spectrum, there are nine localized states.  (c) Eigenstates along the dashed line ($t = 0$) in (a). The localized states, excluding the central zero-energy (dark) state, are doubly degenerate. Probability distributions for typical states are demonstrated in (i)-(iii), which exhibit hybridization of topological corner modes and interface states.
}
\end{figure*}

We present the topological pumping in accordance with the principle of CTAP. In comparison to the conventional adiabatic passage for a three-level system, there are nine topological states involved in our coherent tunneling scheme [see Fig. 2(a) and (c)]. For simplicity, we set $w_x=w_y=1$ as the unit of energy. The other hopping rates are varied following a Gaussian function 
\begin{equation}\label{eq_Gaussian}
\begin{aligned}
	v_x(t)=v_y(t)=\Omega_m \exp\left[-\frac{(t-\frac{\delta}{2})^2}{\lambda^2}\right],\\
	v_x'(t)=v_y(t)'=\Omega_m \exp\left[-\frac{(t+\frac{\delta}{2})^2}{\lambda^2}\right],
\end{aligned}
\end{equation} 
which results in two temporally offset bell-shaped pulses, as shown in Fig. 2(b). This configuration is essential for realizing CTAP, where the excitation transfer is mediated through a dark state \cite{menchon2016spatial, vitanov2017stimulated}. The maximum amplitude is set to $\Omega_m=0.9$, the width is controlled by $\lambda=150$ and we have set a delay constant $\delta = \lambda/3$. Since the hopping rates share the same pattern along $\hat x$ and $\hat y$ directions, we set $M_x=M_y\equiv M$ and $N_x=N_y\equiv N$. The normalization coefficients of the nine topological states in Table \ref{table1} can be easily obtained, yielding 
\begin{equation}\label{N1}
\mathcal{N}_{tl}=\frac{1-M^{2}}{1-M^{L}},
\end{equation}
\begin{equation}\label{N2,4}
\mathcal{N}_{t}=\mathcal{N}_{l}=\frac{1}{\sqrt{\left(\frac{1-M^{L}}{1-M^{2}}\right) \cdot\left(\frac{1-M^{L}}{1-M^{2}}+\frac{1-N^{L}}{1-N^{2}}-1\right)}},
\end{equation}
\begin{equation}\label{N3,7}
\mathcal{N}_{tr}=\mathcal{N}_{bl}=\frac{1}{\sqrt{\left(\frac{1-M^{L}}{1-M^{2}}\right)\left(\frac{1-N^{L}}{1-N^{2}}\right)}},
\end{equation}
\begin{equation}\label{N5}
\mathcal{N}_{c}=\frac{1}{\left(\frac{1-M^{L}}{1-M^{2}}+\frac{1-N^{L}}{1-N^{2}}-1\right)},
\end{equation}
\begin{equation}\label{N6,8}
\mathcal{N}_{r}=\mathcal{N}_{b}=\frac{1}{\sqrt{\left(\frac{1-N^{L}}{1-N^{2}}\right) \cdot\left(\frac{1-N^{L}}{1-N^{2}}+\frac{1-M^{L}}{1-M^{2}}-1\right)}},
\end{equation}
\begin{equation}\label{N9}
\mathcal{N}_{br}=\frac{1-N^{2}}{1-N^{L}}.
\end{equation}

\begin{figure*}[tbp]
    \includegraphics[width=17cm]{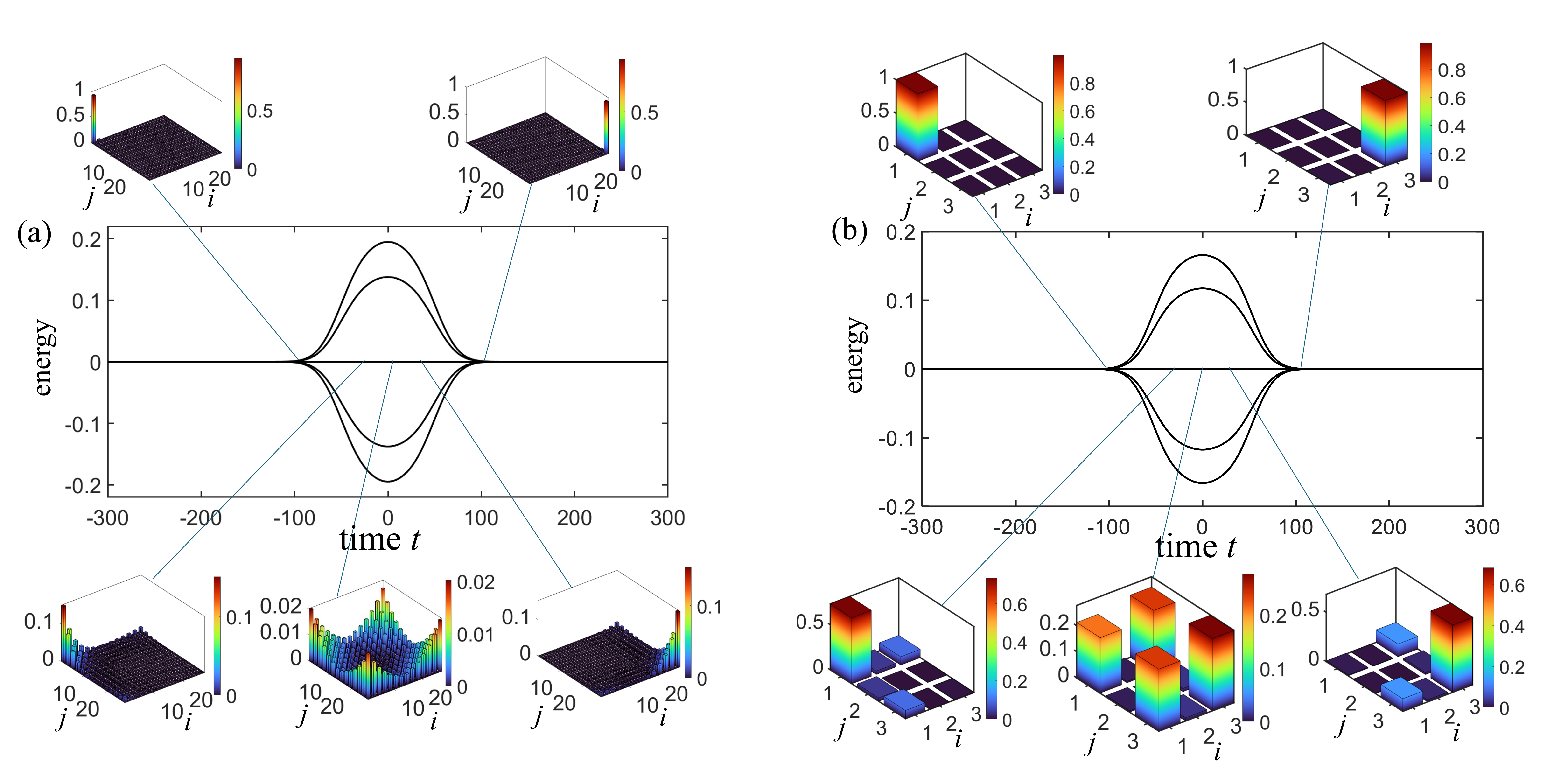}
    \caption{
     (a) Detailed behavior of the eight localized states shown in Fig. 2. Spatial probability distributions of the non-degenerate dark state are illustrated at representative times $t = -100, -25, 0, 25, 100$. The variation of the dark state clearly exhibits a smooth transition from the upper-left corner to the lower-right counterpart in the adiabatic pumping scheme.     (b) Eigenstates of the effective nine-level model. The probability distributions at the same times are depicted. Here $n=3(j-1)+i$ represents the state index of the nine levels. The distribution profiles closely match those of the full model, confirming that the effective CTAP description accurately captures the topological pumping mechanism.
    }
\end{figure*}

We consider an open boundary system with lattice size $27\times27$. The instantaneous spectrum of the time-dependent Hamiltonian $H(t)$ is illustrated in Fig. 2(a). An energy gap exists throughout the time evolution, since the system remains in the topological phase with $v_{x(y)},v_{x(y)'}<w_{x(y)}$. Nine topological states appear within this gap. For $t\to\pm\infty$, the system is in the flat-band limit, these states are degenerate at zero energy. For $t\sim 0$, the degeneracy is lifted, forming a symmetric structure: a dark state pinned at zero energy and four degenerate pairs at positive and negative energies. At $t=0$, typical density distributions for these states are shown in Fig. 2(c). At this point, the dark state is a linear combination of the four corner states ($|TL\rangle,|TR\rangle,|BL\rangle,|BR\rangle$). 

As demonstrated in Fig. 3(a), we observe that the zero-energy eigen-state, which corresponds to the dark state, is localized at the top-left corner $(i,j)=(1,1)$ as $t\to-\infty$, whereas it is localized at the bottom-right corner $(i,j)=(2L-1,2L-1)$ as $t\to+\infty$. This enables robust pumping of a corner state from $|TL\rangle$ to $|BR\rangle$ by adiabatically evolving the dark state.
\subsection{An effective multi-level description}
To gain insight into the dynamics of the pumping scheme, especially in the regime dominated by the in-gap modes, we proceed to construct an effective nine-state model. For this purpose, we focus on the subspace spanned by the localized eigenstates listed in Table \ref{table1}. Within this subspace, a quantum state can be expanded in terms of such topological states. 
\begin{equation}
|\psi(t)\rangle \simeq \sum_{n=1}^9 a_n(t) |n\rangle,
\end{equation}
where the state vectors $|n\rangle$ with $n=1,2,3,\cdots,9$ correspond to $|TL\rangle$, $|T\rangle$, $|TR\rangle$, $\cdots$, $\cdots$, $|BR\rangle$, respectively. The evolution of the amplitudes $a_n$ is governed by 
\begin{equation}
i \frac{d}{dt}
\begin{pmatrix}
a_1 \\
a_2 \\
\vdots \\
a_9 \\

\end{pmatrix}
=
H_{\text{eff}}(t)
\begin{pmatrix}
a_1 \\
a_2 \\
\vdots \\
a_9 \\
\end{pmatrix}.
\end{equation}
Here the effective Hamiltonian $H_{\text{eff}}$ can be obtained in a matrix form as:
\begin{equation}
	\left(\begin{array}{ccccccccc}
          0 & \Omega_{1,2}  & 0 & \Omega_{1,4} & 0 & 0 & 0 & 0 & 0 \\
         \Omega_{1,2}  &  0 & \Omega_{2,3} & 0 & \Omega_{2,5} & 0 & 0 & 0 & 0 \\
          0 & \Omega_{2,3}  & 0 & 0 & 0 & \Omega_{3,6} & 0 & 0 & 0 \\
         \Omega_{1,4} &  0 & 0 & 0 & \Omega_{4,5} & 0 & \Omega_{4,7} & 0 & 0 \\
          0 & \Omega_{2,5}  & 0 & \Omega_{4,5} & 0 & \Omega_{5,6} & 0 & \Omega_{5,8} &  0\\
          0 &  0 & \Omega_{3,6} & 0 & \Omega_{5,6} & 0 & 0 & 0 & \Omega_{6,9} \\
          0 &  0 & 0 & \Omega_{4,7} & 0 & 0 & 0 & \Omega_{7,8} & 0\\
          0 &  0 & 0 & 0 & \Omega_{5,8} & 0 & \Omega_{7,8} & 0 & \Omega_{8,9} \\
          0 &  0 & 0 & 0 & 0 & \Omega_{6,9} & 0 & \Omega_{8,9} & 0 \\
          \end{array}\right)\nonumber
\end{equation}
Here, the coupling strengths stem from the overlap between two topological states upon the action of the Hamiltonian (\ref{eq_H}) \cite{longhi2019topological}. For instance, the coupling between $|TL\rangle$ and $|T\rangle$ is determined by $\Omega_{1,2}=\langle T|H|TL\rangle$. Straightforwardly, we obtain
\begin{equation}
\begin{aligned}
	\Omega_{1,2}=v_x \cdot M^{\left(\frac{L}{2}-1\right)} \cdot\left[\frac{1-M^{L}}{1-M^{2}}\right] \cdot \mathcal N_{tl} \cdot \mathcal N_{t}, \\
	\Omega_{2,3}=v_x' \cdot N^{\left(\frac{L}{2}-1\right)} \cdot\left[\frac{1-N^{L}}{1-N^{2}}\right] \cdot \mathcal N_{t} \cdot \mathcal N_{tr}, \\
\end{aligned}
\end{equation}
and 
\begin{equation}
	\begin{aligned}
		\Omega_{1,2}=-\Omega_{1,4}=\Omega_{4,5}=\Omega_{2,5}=\Omega_{7,8}=-\Omega_{3,6}, \\
		\Omega_{2,3}=-\Omega_{4,7}=\Omega_{5,6}=\Omega_{5,8}=\Omega_{8,9}=-\Omega_{6,9}.
	\end{aligned}
\end{equation}
Diagonalizing $H_\text{eff}$ yields four pairs of degenerate eigenstates with eigenvalues 
\begin{equation}
E_{ \pm}^{(1)}= \pm \sqrt{\Omega_{1,2}^{2}+\Omega_{2,3}^{2}}\quad E_{ \pm}^{(2)}= \pm \sqrt{2} \sqrt{\Omega_{1,2}^{2}+\Omega_{2,3}^{2}},\nonumber
\end{equation}
as well as a zero-energy eigenstate specifically given by
\begin{equation}
|D\rangle = 
\left(
\begin{array}{c}
\dfrac{\Omega_{2,3}^2}{\Omega_{1,2}^2 + \Omega_{2,3}^2} \\
0 \\
-\dfrac{\Omega_{1,2} \Omega_{2,3}}{\Omega_{1,2}^2 + \Omega_{2,3}^2} \\
0 \\
0 \\
0 \\
-\dfrac{\Omega_{1,2} \Omega_{2,3}}{\Omega_{1,2}^2 + \Omega_{2,3}^2} \\
0 \\
\dfrac{\Omega_{1,2}^2}{\Omega_{1,2}^2 + \Omega_{2,3}^2}
\end{array}
\right).
\end{equation}
Clearly, $|D\rangle$ is the dark state without populations on the interface states and the center state. According to the mechanism of CTAP or STIRAP, as the hopping rates follow the variations in Eqs. (\ref{eq_Gaussian}), we have $v_x/v_x'\to 0$ as $t\to -\infty$, where $\Omega_{1,2}/\Omega_{2,3}\to 0$ and $D\to|TL\rangle$. Conversely, $v_x'/v_x\to 0$ as $t\to +\infty$, in which case $\Omega_{2,3}/\Omega_{1,2}\to 0$ and $D\to|BR\rangle$. In contrast to a three-level system, the dark state involves finite distribution in $|TR\rangle$ and $|BL\rangle$, which can be taken as an intermediate process. In fact, we have $|a_1|^2+|a_9|^2\geq|a_3|^2+|a_7|^2$ even for $t\sim 0$. Therefore, robust corner state transfer can be achieved by engineering the dark state with adiabatic evolution. 

Figure 3 demonstrates the excellent agreement between the complete and effective models. The spectra and distributions are obtained by diagonalizing the full Hamiltonian and the effective Hamiltonian matrix, with hopping rates following the modulation in Fig. 2(b). We have depicted the probability distribution profiles of the evolving dark state at selected times $t=-100,-25,0,25,100$. The amplitudes associated with interface modes remain negligible during the whole evolution. These results indicate that the constructed nine-level model successfully captures the essential dynamics of the system within the subspace of topological states and provides an accurate description of the corner-to-corner transport under the CTAP scheme.

\subsection{Numerical results}
\begin{figure}[tbp]
    \includegraphics[width=8.5cm]{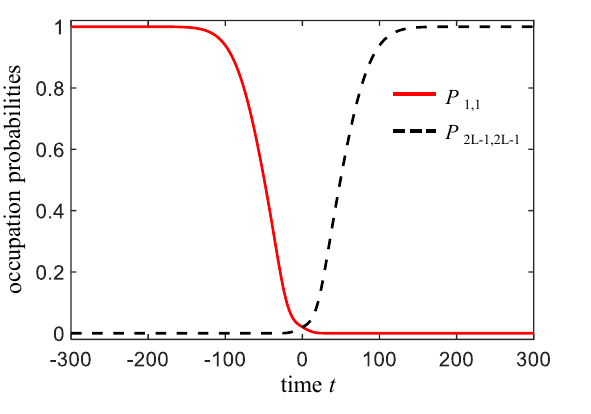}
    \caption{
    Evolution of the occupation probabilities at the upper-left corner site $(i,j)=(1,1)$ and the lower-right corner site $(2L-1,2L-1)$. The total modulation time is $T=600$. The system is initialized with a fully localized state at $(1,1)$ and evolves according to the variation of hopping rates shown in Fig. 2. 
    }
\end{figure}
\begin{figure}[tbp]
    \includegraphics[width=8.8cm]{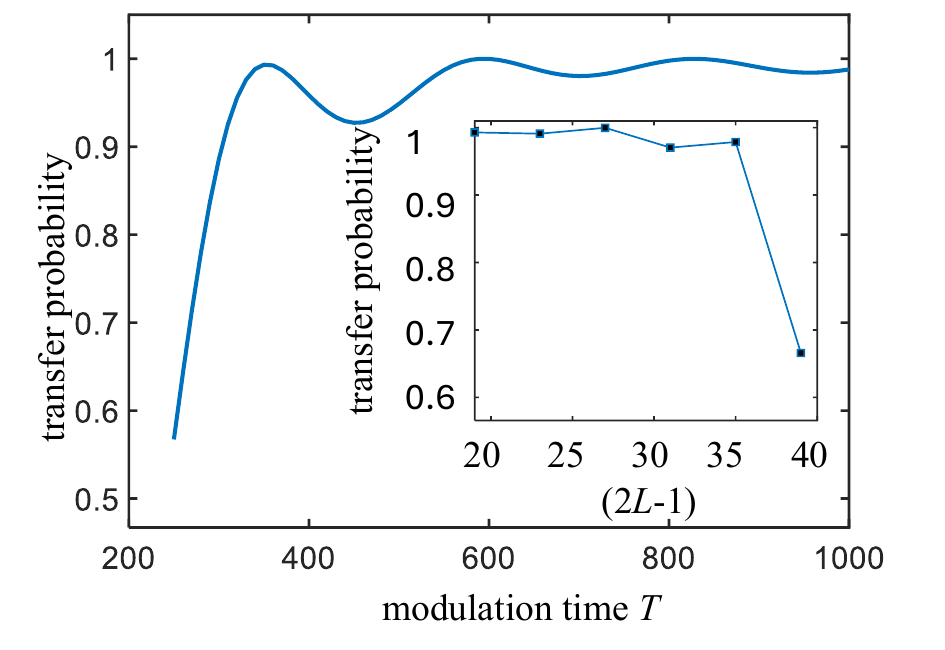}
    \caption{
     Transfer probability $P_{{2L-1},{2L-1}}(T)$ as a function of the total modulation time $T$. When $T \gtrsim 360$, the pumping achieves a transfer probability exceeding 90\%. The inset depicts the behavior of the transfer probability versus system size for a fixed modulation time $T = 600$.
    }
\end{figure}
We start with a corner state $\hat c_{1,1}^\dag|0\rangle$ which is fully localized on the top-left vertex. The propagation of the initial state is simulated by solving the time-dependent Schr\"odinger equation $i\partial_t|\psi(t)\rangle=\hat H(t)|\psi(t)\rangle$. The system parameters are the same as those used in Fig. 2 and 3. During the time evolution, we calculate the occupation probabilities $P_{i,j}$ at sites $(i,j)=(1,1)$ and $(2L-1,2L-1)$. Results for an evolution time $T=600$ are shown in Fig. 4. We observe a high-fidelity corner state transfer with transfer probability $P_{2L-1,2L-1}=99.98$\%. Besides, the final state fidelity as a function of the evolution time $T$ is depicted in Fig. 5. For each simulation, hopping rates are tuned according to Eqs. (\ref{eq_Gaussian}) with $\lambda=3T/10$, $\delta=\lambda/3$ and $\Omega_m=0.9$. The system enters the adiabatic regime when $T\gtrsim 360$ with transfer probability exceeding 90\%. For a fixed duration $T=600$, the transfer probability versus lattice size is shown in the inset of Fig. 5. As the system size increase, the fidelity collapses, indicating a rapid breakdown of the adiabatic condition \cite{vitanov2017stimulated}
\begin{equation}
A \equiv \int_{-\infty}^{\infty} dt \, \sqrt{\Omega_{1,2}(t)^2 + \Omega_{2,3}(t)^2} \gg \frac{\pi}{2}.
\end{equation}
This occurs since $\Omega_{1,2}$ and $\Omega_{2,3}$ decrease nearly exponentially with increasing $L$ \cite{longhi2019topological}. Therefore, maintaining adiabaticity in larger systems necessitates a longer evolution time. 

\section{comparison with a Thouless pumping scheme}\label{sec4}
\begin{figure*}[tbp]
    \includegraphics[width=16cm]{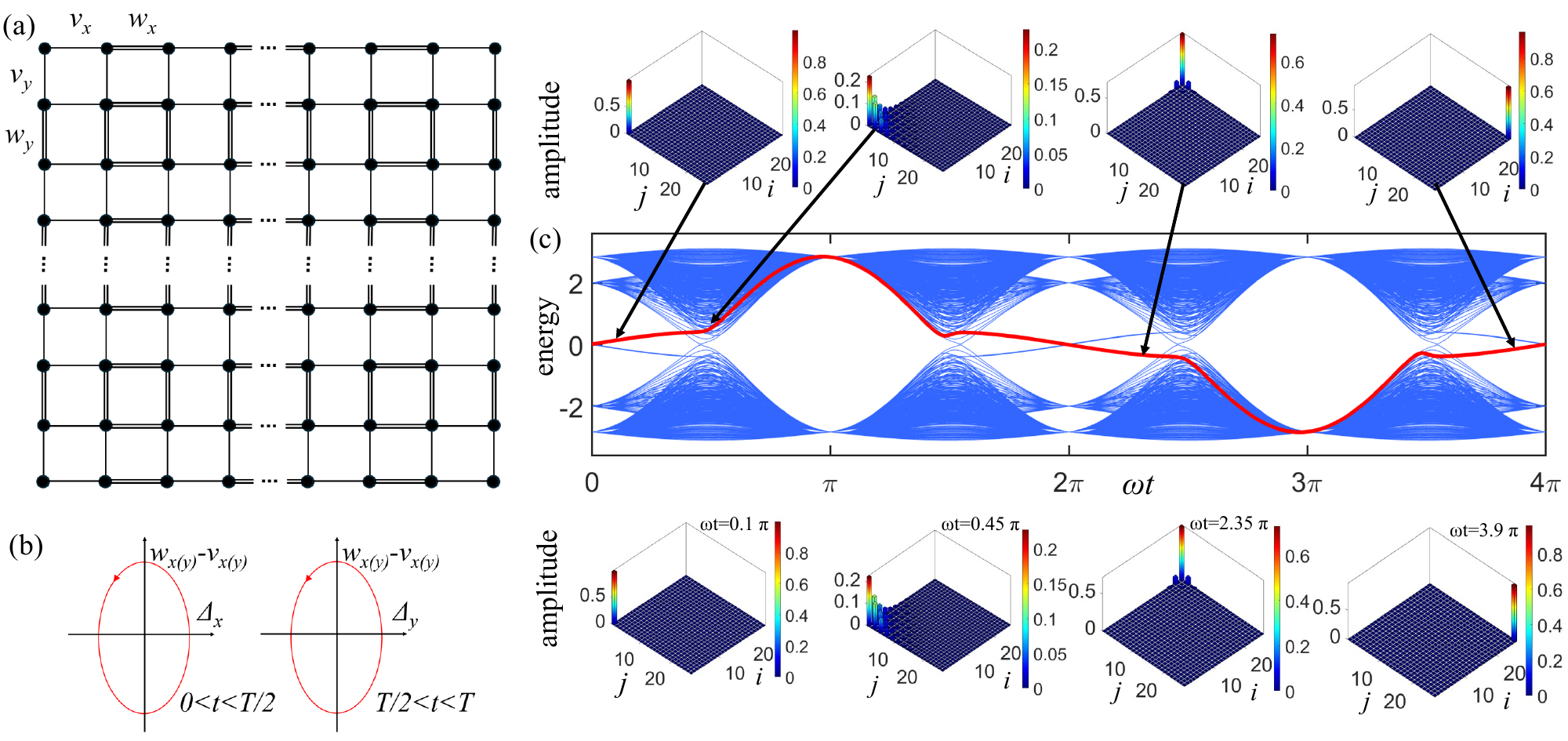}
\caption{
(a) Schematic diagram of the lattice structure, characterized by hopping rates $v_{x(y)},w_{x(y)}$. Staggered potential $\Delta_x$ or $\Delta_y$ is applied according to Eq. (\ref{eq_potential}), forming a two-dimensional extension of the Rice-Mele model. (b) Two-stage adiabatic path of parameters. We have set $v_x=v_y$ and $w_x=w_y$. (c) Energy spectra with respect to $\omega t$. The red curve highlights the variation of the initial corner modes. Spacial probability distributions from exact diagonalizations are presented in the upper panel, respectively for $\omega t = 0.1\pi,0.45\pi,2.35\pi,3.9\pi$. During the evolution starting from a fully localized corner state at $(i,j)=(1,1)$, the distributions of instantaneous wavefunction are shown in the lower panel, exhibiting close agreement with the exact diagonalization results.
}

\end{figure*}

The CTAP-based protocol employs a counter-intuitive modulation sequence that separates the topological dark state from other excitations, leading to a near-complete transfer of population without significantly populating the interface states. Consequently, the inherent adiabaticity confers exceptional robustness against variations in hopping rates and detunings. This makes the present topological pumping scheme distinct from direct resonant pumping \cite{st2017lasing}, Thouless pumping \cite{thouless1983quantization, citro2023thouless} or other schemes. As a comparison, we present the performance of a 2D version of the Rice-Mele scheme in corner state pumping. 

The model is based on a 2D SSH lattice under open boundary condition, as demonstrated in Fig. 6(a). In addition to the alternating tunneling strengths $v_x$, $w_x$ and $v_y$, $w_y$, we have included a staggered on-site potential 
\begin{equation}\label{eq_potential}
	\hat H_\Delta = \sum_{i,j=1}^{2L}\left[(-1)^{i}\Delta_x+(-1)^{j}\Delta_y\right]\hat n_{i,j},
\end{equation}
which gives rise to a Bloch Hamiltonian 
\begin{equation}
	\hat H_k'=\hat H_k+\Delta_x I\otimes\sigma_z+\Delta_y\sigma_z\otimes I.
\end{equation}
$\hat H_k$ has spectrum 
\begin{equation}
	E=\pm\sqrt{\sum_{n=x,y}(v_n^2+w_n^2+v_nw_n\cos k_n)}
\end{equation}
which closes at critical point $v_x=w_x$, $v_y=w_y$ (with $k_x=k_y=\pi$). For a 1D system, the Rice-Mele pump is realized by modulating the tunneling strength and on-site potential to form a closed loop in parameter space that encircles the critical point \cite{citro2023thouless}. Based on this strategy, we consider a two-stage protocol, where each stage consists of a cyclic modulation of a distinct parameter set, as shown in Fig. 6(b). Specifically, we take 
\begin{equation}
	\begin{aligned}
		v_x&=v_y=t_0\{1-\cos[\omega t]\},\\
		w_x&=w_y=t_0\{1+\cos[\omega t]\},\\
		\Delta_x&=\delta_0\sin[\omega t\Theta(T/2-t)],\\
		\Delta_y&=\delta_0\sin[\omega t\Theta(t-T/2)],
	\end{aligned}
\end{equation}
with $\omega=4\pi/T$, $t\in[0,T]$ and $\Theta$ the Heaviside step function. In this case, we set $t_0=1$ as the unit of energy .
\begin{figure}[tbp]
    \includegraphics[width=8.5cm]{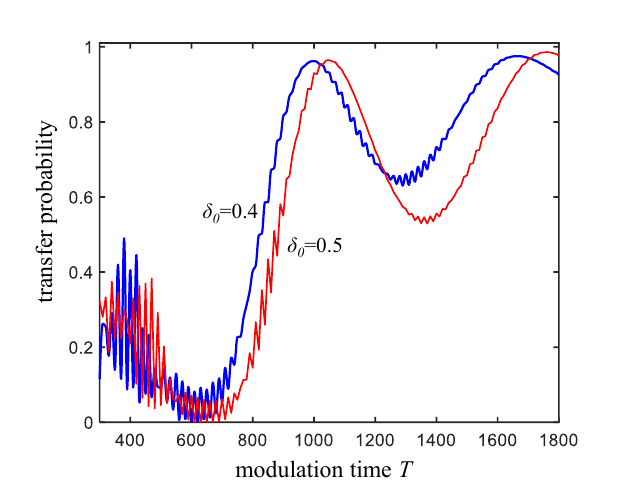}
   \caption{
 Final-state transfer probability as a function of total evolution time $T$ for a $2D$ SSH lattice with $2L = 26$.
}
\end{figure}

The instantaneous eigenvalues as a function of $\omega t$ for a system with lattice size $(2L\times 2L)=(26\times26)$ are plotted in Fig. 6(c). Distributions for typical eigenvectors are shown in the upper panel. At $t=0$, the corner states are four-fold degenerate, corresponding to the single-site excitation on each vertex. During the modulation, the degeneracy is lifted to a two-fold one. In the first stage, due to the modulation of staggered on-site potential $\pm\Delta_x$ along $\hat x$ direction, the upper-left corner mode is transformed into the upper-right corner state at $t=T/2$. Analogously, the second stage carries out a $\hat y$ direction pumping, which ends up with a localized state at $(i,j)=(2L,2L)$. We take the initial state as the one fully localized at site $(i,j)=(1,1)$, which is expected to follow the trajectory of the associated state upon adiabatic evolution. The lower panel in Fig. 6(c) demonstrates the instantaneous wavefunction distribution at particular time $t=0.1\pi,0.45\pi,2.35\pi,3.9\pi$ (with $T=1000$, $\delta_0=0.4$) in real-time evolution. This simulation shows close agreement with the results of exact diagonalization. Note that the in-gap corner modes merge into the bulk, therefore, an extremely slow modulation is necessary to maintain adiabaticity. The variation of transfer probability $P_{2L,2L}$ with respect to modulation time $T$ is plotted in Fig. 7. The low fidelity for $T\lesssim900$ indicates pronounced nonadiabatic effect. The strong oscillation may be caused by Rabi-like oscillations among the corner and the bulk states. For $\delta_0 = 0.4$ ($ 0.5$), the transfer probability reaches 0.962 (0.964) at $T=1000$ (1050). However, stable pumping with high level of fidelity requires modulation time even larger than 1800. This result stands in sharp contrast to the performance of previous CTAP scheme. 

\section{summary}\label{sec5}
The active manipulation of topologically protected states, encompassing not just edge states but also localized corner modes and tailored interface states, represents a pivotal frontier in modern physics. The significance lies in the unique confluence of topological robustness and the capacity for quantum control. Among quantum control techniques for population transfer, CTAP and STIRAP stand apart due to their unique robustness and efficiency. In this work, we presented a protocol for corner state transfer by synthesizing the concepts of higher-order topology and coherent tunneling by adiabatic passage. The scheme is implemented in a 2D SSH model composed of four sub-blocks with distinct topological features. Such a system sustains topological dark state which can be manipulated via integrated modulation. We have demonstrated that the adiabatic pumping scheme facilitates the robust transfer of a localized topological excitation between two distal corner sites of the lattice. We further formulated a nine-level model that captures the essential physics of the underlying mechanism. In comparison with a two-stage Thouless pumping, the present protocol offers enhanced performance in both fidelity and efficiency. 

This work establishes a scalable framework for robust excitation transfer of topologically protected states in a two-dimensional lattice, which can be directly extended to higher-dimensional quantum systems. The generality of this scheme suggests potential implementations in diverse platforms, including superconducting circuits \cite{deng2024high}, photonic lattices \cite{cerjan2020thouless} and cold atom systems \cite{lohse2016thouless}. This methodology provides a promising foundation for future quantum information processing and quantum control that leverage topological dark state. The present scheme could be further enhanced by integrating techniques developed for accelerating adiabatic protocols. Well-established methods including shortcuts to adiabaticity \cite{longhi2019topological, baksic2016speeding, guery2019shortcuts, d2020fast}, optimal control theory \cite{caneva2009optimal, machnes2011comparing, palaiodimopoulos2021fast} and composite pulse sequences \cite{genov2017arbitrarily} offer direct pathways to suppress non-adiabatic losses and achieve high-fidelity state transfer on significantly shorter timescales.

\section{Acknowledgements}
This work is supported by NSFC (Grant No. 12304179) and Sichuan Normal University College Students' Innovation and Entrepreneurship Training Program(Grant No. 202510636005).

\bibliography{TP.bib}

\begin{thebibliography}{10}

\bibitem{thouless1982quantized}
D.~J. Thouless, M.~Kohmoto, M.~P. Nightingale, and M.~den Nijs,
\newblock Physical Review Letters {\bf 49}, 405 (1982).

\bibitem{schnyder2008classification}
A.~P. Schnyder, S.~Ryu, A.~Furusaki, and A.~W. Ludwig,
\newblock Physical Review B {\bf 78}, 195125 (2008).

\bibitem{hasan2010colloquium}
M.~Z. Hasan and C.~L. Kane,
\newblock Reviews of Modern Physics {\bf 82}, 3045 (2010).

\bibitem{ryu2010topological}
S.~Ryu, A.~P. Schnyder, A.~Furusaki, and A.~W. Ludwig,
\newblock New Journal of Physics {\bf 12}, 065010 (2010).

\bibitem{qi2011topological}
X.-L. Qi and S.-C. Zhang,
\newblock Reviews of Modern Physics {\bf 83}, 1057 (2011).

\bibitem{bansil2016colloquium}
A.~Bansil, H.~Lin, and T.~Das,
\newblock Reviews of Modern Physics {\bf 88}, 021004 (2016).

\bibitem{su1979solitons}
W.-P. Su, J.~R. Schrieffer, and A.~J. Heeger,
\newblock Physical Review Letters {\bf 42}, 1698 (1979).

\bibitem{creutz1999end}
M.~Creutz,
\newblock Physical Review Letters {\bf 83}, 2636 (1999).

\bibitem{haldane1988model}
F.~D.~M. Haldane,
\newblock Physical Review Letters {\bf 61}, 2015 (1988).

\bibitem{delplace2011zak}
P.~Delplace, D.~Ullmo, and G.~Montambaux,
\newblock Physical Review B {\bf 84}, 195452 (2011).

\bibitem{grusdt2013topological}
F.~Grusdt, M.~H{\"o}ning, and M.~Fleischhauer,
\newblock Physical Review Letters {\bf 110}, 260405 (2013).

\bibitem{nakajima2016topological}
S.~Nakajima, T.~Tomita, S.~Taie, T.~Ichinose, H.~Ozawa, L.~Wang, M.~Troyer, and
  Y.~Takahashi,
\newblock Nature Physics {\bf 12}, 296 (2016).

\bibitem{sun2017quantum}
N.~Sun and L.-K. Lim,
\newblock Physical Review B {\bf 96}, 035139 (2017).

\bibitem{junemann2017exploring}
J.~J{\"u}nemann, A.~Piga, S.-J. Ran, M.~Lewenstein, M.~Rizzi, and
  A.~Berm{\'u}dez,
\newblock Physical Review X {\bf 7}, 031057 (2017).

\bibitem{cooper2019topological}
N.~Cooper, J.~Dalibard, and I.~Spielman,
\newblock Reviews of Modern Physics {\bf 91}, 015005 (2019).

\bibitem{thouless1983quantization}
D.~Thouless,
\newblock Physical Review B {\bf 27}, 6083 (1983).

\bibitem{ke2017multiparticle}
Y.~Ke, X.~Qin, Y.~S. Kivshar, and C.~Lee,
\newblock Physical Review A {\bf 95}, 063630 (2017).

\bibitem{fedorova2020observation}
Z.~Fedorova, H.~Qiu, S.~Linden, and J.~Kroha,
\newblock Nature Communications {\bf 11}, 3758 (2020).

\bibitem{citro2023thouless}
R.~Citro and M.~Aidelsburger,
\newblock Nature Reviews Physics {\bf 5}, 87 (2023).

\bibitem{lv2025exploring}
Y.-J. Lv, Y.~Peng, Y.-K. Liu, and Y.~Zheng,
\newblock Physical Review A {\bf 111}, 013311 (2025).

\bibitem{rice1982elementary}
M.~Rice and E.~Mele,
\newblock Physical Review Letters {\bf 49}, 1455 (1982).

\bibitem{xiao2010berry}
D.~Xiao, M.-C. Chang, and Q.~Niu,
\newblock Reviews of Modern Physics {\bf 82}, 1959 (2010).

\bibitem{wang2013topological}
L.~Wang, M.~Troyer, and X.~Dai,
\newblock Physical Review Letters {\bf 111}, 026802 (2013).

\bibitem{niu1984quantised}
Q.~Niu and D.~Thouless,
\newblock Journal of Physics A: Mathematical and General {\bf 17}, 2453 (1984).

\bibitem{kraus2012topological}
Y.~E. Kraus, Y.~Lahini, Z.~Ringel, M.~Verbin, and O.~Zilberberg,
\newblock Physical Review Letters {\bf 109}, 106402 (2012).

\bibitem{lohse2016thouless}
M.~Lohse, C.~Schweizer, O.~Zilberberg, M.~Aidelsburger, and I.~Bloch,
\newblock Nature Physics {\bf 12}, 350 (2016).

\bibitem{xu2017topological}
Z.~Xu, Y.~Zhang, and S.~Chen,
\newblock Physical Review A {\bf 96}, 013606 (2017).

\bibitem{longhi2019topological}
S.~Longhi,
\newblock Physical Review B {\bf 99}, 155150 (2019).

\bibitem{zurita2020topology}
J.~Zurita, C.~E. Creffield, and G.~Platero,
\newblock Advanced Quantum Technologies {\bf 3}, 1900105 (2020).

\bibitem{minguzzi2022topological}
J.~Minguzzi, Z.~Zhu, K.~Sandholzer, A.-S. Walter, K.~Viebahn, and T.~Esslinger,
\newblock Physical Review Letters {\bf 129}, 053201 (2022).

\bibitem{connolly2013gigahertz}
M.~Connolly {\em et~al.},
\newblock Nature Nanotechnology {\bf 8}, 417 (2013).

\bibitem{kandel2021adiabatic}
Y.~P. Kandel, H.~Qiao, S.~Fallahi, G.~C. Gardner, M.~J. Manfra, and J.~M.
  Nichol,
\newblock Nature Communications {\bf 12}, 2156 (2021).

\bibitem{yatsugi2022observation}
K.~Yatsugi, T.~Yoshida, T.~Mizoguchi, Y.~Kuno, H.~Iizuka, Y.~Tadokoro, and
  Y.~Hatsugai,
\newblock Communications Physics {\bf 5}, 180 (2022).

\bibitem{ryu2002topological}
S.~Ryu and Y.~Hatsugai,
\newblock Physical Review Letters {\bf 89}, 077002 (2002).

\bibitem{nayak2008non}
C.~Nayak, S.~H. Simon, A.~Stern, M.~Freedman, and S.~Das~Sarma,
\newblock Reviews of Modern Physics {\bf 80}, 1083 (2008).

\bibitem{gorg2019realization}
F.~G{\"o}rg, K.~Sandholzer, J.~Minguzzi, R.~Desbuquois, M.~Messer, and
  T.~Esslinger,
\newblock Nature Physics {\bf 15}, 1161 (2019).

\bibitem{de2019observation}
S.~De~L{\'e}s{\'e}leuc, V.~Lienhard, P.~Scholl, D.~Barredo, S.~Weber, N.~Lang,
  H.~P. B{\"u}chler, T.~Lahaye, and A.~Browaeys,
\newblock Science {\bf 365}, 775 (2019).

\bibitem{mellnik2014spin}
A.~R. Mellnik {\em et~al.},
\newblock Nature {\bf 511}, 449 (2014).

\bibitem{han2018quantum}
W.~Han, Y.~Otani, and S.~Maekawa,
\newblock npj Quantum Materials {\bf 3}, 27 (2018).

\bibitem{gaubatz1990population}
U.~Gaubatz, P.~Rudecki, S.~Schiemann, and K.~Bergmann,
\newblock The Journal of Chemical Physics {\bf 92}, 5363 (1990).

\bibitem{bergmann1998coherent}
K.~Bergmann, H.~Theuer, and B.~Shore,
\newblock Reviews of Modern Physics {\bf 70}, 1003 (1998).

\bibitem{menchon2016spatial}
R.~Menchon-Enrich, A.~Benseny, V.~Ahufinger, A.~D. Greentree, T.~Busch, and
  J.~Mompart,
\newblock Reports on Progress in Physics {\bf 79}, 074401 (2016).

\bibitem{vitanov2017stimulated}
N.~V. Vitanov, A.~A. Rangelov, B.~W. Shore, and K.~Bergmann,
\newblock Reviews of Modern Physics {\bf 89}, 015006 (2017).

\bibitem{leroux2018non}
F.~Leroux, K.~Pandey, R.~Rehbi, F.~Chevy, C.~Miniatura, B.~Gr{\'e}maud, and
  D.~Wilkowski,
\newblock Nature Communications {\bf 9}, 3580 (2018).

\bibitem{benalcazar2017quantized}
W.~A. Benalcazar, B.~A. Bernevig, and T.~L. Hughes,
\newblock Science {\bf 357}, 61 (2017).

\bibitem{benalcazar2017electric}
W.~A. Benalcazar, B.~A. Bernevig, and T.~L. Hughes,
\newblock Physical Review B {\bf 96}, 245115 (2017).

\bibitem{li2020topological}
C.-A. Li and S.-S. Wu,
\newblock Physical Review B {\bf 101}, 195309 (2020).

\bibitem{serra2018observation}
M.~Serra-Garcia, V.~Peri, R.~S{\"u}sstrunk, O.~R. Bilal, T.~Larsen, L.~G.
  Villanueva, and S.~D. Huber,
\newblock Nature {\bf 555}, 342 (2018).

\bibitem{yang2020type}
Y.-B. Yang, K.~Li, L.-M. Duan, and Y.~Xu,
\newblock Physical Review Research {\bf 2}, 033029 (2020).

\bibitem{vergara2024emerging}
P.~Vergara, G.~S{\'a}ez, M.~Castro, S.~Allende, and {\'A}.~S. N{\'u}{\~n}ez,
\newblock npj 2D Materials and Applications {\bf 8}, 41 (2024).

\bibitem{xie2018second}
B.-Y. Xie, H.-F. Wang, H.-X. Wang, X.-Y. Zhu, J.-H. Jiang, M.-H. Lu, and Y.-F.
  Chen,
\newblock Physical Review B {\bf 98}, 205147 (2018).

\bibitem{obana2019topological}
D.~Obana, F.~Liu, and K.~Wakabayashi,
\newblock Physical Review B {\bf 100}, 075437 (2019).

\bibitem{kim2020topological}
M.~Kim and J.~Rho,
\newblock Nanophotonics {\bf 9}, 3227 (2020).

\bibitem{yang2021robust}
Z.-Z. Yang, A.-Y. Guan, W.-J. Yang, X.-Y. Zou, and J.-C. Cheng,
\newblock arXiv:2102.12234  (2021).

\bibitem{li2022topological}
C.-A. Li,
\newblock Frontiers in Physics {\bf 10}, 861242 (2022).

\bibitem{wei2023topological}
M.-S. Wei, M.-J. Liao, C.~Wang, C.~Zhu, Y.~Yang, and J.~Xu,
\newblock Optics Express {\bf 31}, 3427 (2023).

\bibitem{mittal2019photonic}
S.~Mittal, V.~V. Orre, G.~Zhu, M.~A. Gorlach, A.~Poddubny, and M.~Hafezi,
\newblock Nature Photonics {\bf 13}, 692 (2019).

\bibitem{lu2014topological}
L.~Lu, J.~D. Joannopoulos, and M.~Solja{\v{c}}i{\'c},
\newblock Nature Photonics {\bf 8}, 821 (2014).

\bibitem{el2019corner}
A.~El~Hassan, F.~K. Kunst, A.~Moritz, G.~Andler, E.~J. Bergholtz, and
  M.~Bourennane,
\newblock Nature Photonics {\bf 13}, 697 (2019).

\bibitem{deng2024high}
C.-L. Deng {\em et~al.},
\newblock Physical Review Letters {\bf 133}, 140402 (2024).

\bibitem{benalcazar2022higher}
W.~A. Benalcazar, J.~Noh, M.~Wang, S.~Huang, K.~P. Chen, and M.~C. Rechtsman,
\newblock Physical Review B {\bf 105}, 195129 (2022).

\bibitem{st2017lasing}
P.~St-Jean, V.~Goblot, E.~Galopin, A.~Lema{\^\i}tre, T.~Ozawa, L.~Le~Gratiet,
  I.~Sagnes, J.~Bloch, and A.~Amo,
\newblock Nature Photonics {\bf 11}, 651 (2017).

\bibitem{cerjan2020thouless}
A.~Cerjan, M.~Wang, S.~Huang, K.~P. Chen, and M.~C. Rechtsman,
\newblock Light: Science \& Applications {\bf 9}, 178 (2020).

\bibitem{baksic2016speeding}
A.~Baksic, H.~Ribeiro, and A.~A. Clerk,
\newblock Physical Review Letters {\bf 116}, 230503 (2016).

\bibitem{guery2019shortcuts}
D.~Gu{\'e}ry-Odelin, A.~Ruschhaupt, A.~Kiely, E.~Torrontegui,
  S.~Mart{\'\i}nez-Garaot, and J.~G. Muga,
\newblock Reviews of Modern Physics {\bf 91}, 045001 (2019).

\bibitem{d2020fast}
F.~M. d'Angelis, F.~A. Pinheiro, D.~Gu{\'e}ry-Odelin, S.~Longhi, and F.~Impens,
\newblock Physical Review Research {\bf 2}, 033475 (2020).

\bibitem{caneva2009optimal}
T.~Caneva, M.~Murphy, T.~Calarco, R.~Fazio, S.~Montangero, V.~Giovannetti, and
  G.~E. Santoro,
\newblock Physical Review Letters {\bf 103}, 240501 (2009).

\bibitem{machnes2011comparing}
S.~Machnes, U.~Sander, S.~J. Glaser, P.~de~Fouquieres, A.~Gruslys, S.~Schirmer,
  and T.~Schulte-Herbr{\"u}ggen,
\newblock Physical Review A {\bf 84}, 022305 (2011).

\bibitem{palaiodimopoulos2021fast}
N.~Palaiodimopoulos, I.~Brouzos, F.~Diakonos, and G.~Theocharis,
\newblock Physical Review A {\bf 103}, 052409 (2021).

\bibitem{genov2017arbitrarily}
G.~T. Genov, D.~Schraft, N.~V. Vitanov, and T.~Halfmann,
\newblock Physical Review Letters {\bf 118}, 133202 (2017).

\end{thebibliography}

\end{document}